\documentclass[aps,prb,twocolumn,amsmath,showpacs,amssymb,superscriptaddress]{revtex4}

\usepackage{amsmath}
\usepackage{graphicx}
\usepackage{amssymb}
\usepackage{dsfont}
\usepackage{color}

\newcommand{\pdag}{{\phantom{\dagger}}}

\begin{document}

\title{Superconducting Gap Renormalization around two Magnetic Impurities: \\ From Shiba to
Andreev Bound States}

\author{Tobias Meng}
\affiliation{Institut f\"ur Theoretische Physik, Technische Universit\"at Dresden, 01062 Dresden, Germany}
\affiliation{Department of Physics, University of Basel, Klingelbergstrasse 82, CH-4056 Basel, Switzerland}

\author{Jelena Klinovaja}
\affiliation{Department of Physics, University of Basel, Klingelbergstrasse 82, CH-4056 Basel, Switzerland}

\author{Silas Hoffman}
\affiliation{Department of Physics, University of Basel, Klingelbergstrasse 82, CH-4056 Basel, Switzerland}

\author{Pascal Simon}
\affiliation{Laboratoire de Physique des Solides, CNRS UMR-8502, Universit\'{e} Paris Sud, 91405 Orsay Cedex, France}

\author{Daniel Loss}
\affiliation{Department of Physics, University of Basel, Klingelbergstrasse 82, CH-4056 Basel, Switzerland}

\begin{abstract}
We study the renormalization of the gap of an $s$-wave superconductor in the presence of two magnetic impurities. For weakly bound Shiba states, we analytically  calculate the part of the gap renormalization that is sensitive to the relative orientation of the two impurity spins. 
For impurities with a strong exchange coupling to the conduction electrons, we solve the gap equation self-consistently by numerics and find that the sub-gap Shiba state turns into a supra-gap Andreev state when the local gap parameter changes sign under the impurities.
\end{abstract}

\maketitle

The study of magnetic impurities in superconductors has a long history. Important effects include the renormalization (reduction) of the gap in the host superconductor by the impurities,\cite{rusinov_69,schlottmann_76} and the induced Yu-Shiba-Rusinov, or  Shiba, bound states.\cite{shiba_68,yu_65,rusinov_68,sakurai_70,yazdani_97,flatte_97,salkola_97,flatte_00,
morr_06,balatsky_06,ji_08,rkky_sc_14,zarand_14,zyuzin_14,Silas} In recent years, renewed interest in chains of magnetic impurities in superconductors was furthermore driven by their potential of hosting Majorana zero modes.
\cite{nadj_13,klinovaja_13,braunecker_13,vazifeh_13, pientka_13, nakosai_13,fins_14,reis_14,nadj_14,meyer_15}

Using scanning tunneling microscopy, the Shiba states induced by individual magnetic impurities on the surface of a superconductor have been shown to be strongly localized at the impurity sites.\cite{yazdani_97,ji_08} This can be understood as a consequence of their sub-gap energy. When this energy is below the Fermi level, the Shiba state is occupied by a single electron, whose spin direction is dictated by the impurity spin.\cite{sakurai_70} Shiba states are to be contrasted with Andreev bound states in extended superconductor-normal-superconductor (S-N-S) heterostructures, or in  heterostructures composed of superconductors with different gaps (S-S'-S). Andreev bound states can be viewed as standing waves of electrons and holes inside the gapless region that result from Andreev reflections on the S-N-S and S-S'-S interfaces.\cite{andreev_abs_66,kulik_70,rowell_73,klapwijk_82}

In this work, we address a scenario intermediate between single impurities and impurity chains, and specifically study the renormalization of the superconducting order parameter in the presence of two magnetic impurities. We quantify the  inter-impurity scattering analytically and discuss how the Shiba states are affected by the gap renormalization. In particular, we observe that sub-gap Shiba states can transmute into supra-gap Andreev states when the impurities are strongly bound to the superconductor, which we show by a simple analytical model and by self-consistent numerics.

The outline of the paper is as follows. In the next section, we introduce our model Hamiltonian for the magnetic impurity in an $s$-wave superconductor. In Sec.~\ref{sec:t_matrix}, we summarize briefly our analytical approach based on the T-matrix formalism. In Sec.~\ref{sec:single_imp}, we apply it to a single classical impurity. In Sec.~\ref{sec:shiba_andreev}, we present a detailed study of the Shiba to Andreev transition as a function of the impurity strength using both analytical and numerical methods. In Sec.~\ref{sec:two_imp},  we consider the two-impurity case, and show the presence of two subsequent quantum phase transitions as a function of the exchange coupling. Finally, in Sec.~\ref{sec:concl}, we present a summary of our results and some perspectives.

\section{Model Hamiltonian}\label{sec:model_ham}
In the analytical part of our study, we analyze two (weakly bound) magnetic impurities in a three-dimensional $s$-wave superconductor. The corresponding Hamiltonian $H=H_{\rm kin} + H_{\rm SC} + H_{\rm imp}$ can be decomposed into the kinetic energy, $H_{\rm kin}$, the BCS mean field superconductivity, $H_{\rm SC}$, and the impurity contribution, $H_{\rm imp}$. If there were no impurities, the system could be modeled by the Hamiltonian $H_0=H_{{\rm kin}} + H_{{\rm SC},0}$, where $H_{{\rm kin}}=\sum_{\textbf{k},\sigma} E_{\bf k}\,c_{ {\bf k}\sigma}^\dagger c_{ {\bf k}\sigma}^\pdag + \rm{H.c.}$ ($E_{\bf k}$ denotes the kinetic energy dispersion), and $H_{{\rm SC},0}=\int d{\bf r} \,\Delta_0  \,c_\uparrow^\dagger({\bf r}) c_\downarrow^\dagger({\bf r}) + \rm{H.c.}$ with a  superconducting order parameter of constant value $\Delta_0$ (chosen to be positive). Here, $c_{ {\bf k}\sigma}$ is the annihilation operator for an electron of three-dimensional momentum $\bf k$ and spin $\sigma$, while $c_{\sigma}^\dagger({\bf r})$ creates a spin $\sigma$ electron at position ${\bf r}$ (in the calculations, we use $\uparrow\equiv +$, $\downarrow \equiv -$). Due to the impurities, however, the superconducting part of the Hamiltonian is promoted from $H_{{\rm SC},0}$ to
\begin{align}
H_{{\rm SC}}=\int d{\bf r} \,\Delta({\bf r}) \,c_\uparrow^\dagger({\bf r}) c_\downarrow^\dagger({\bf r}) + \rm{H.c.}
\end{align}
The order parameter is set by the self-consistent equation $\Delta({\bf r}) = -g\,\langle c_\downarrow({\bf r}) c_\uparrow({\bf r})\rangle_H$, where $g>0$ is the microscopic attractive interaction between electrons in the superconductor, and where the expectation value is taken with respect to the full $H$. The magnetic impurities, finally, are treated as classical spins ${\bf{S}}_i$ residing at positions ${\bf{r}}_i$, and modeled by a purely magnetic~\cite{okabe_83} and point-like scattering potential. Their classical treatment implies that the we restrict our analysis to sufficiently large impurity spins. We address impurities polarized along the $\hat{z}$ axis in spin space, thus covering both parallel and antiparallel impurity spin alignments. 
The corresponding impurity Hamiltonian reads
\begin{align}
H_{\rm imp} = \sum_{i,\sigma} J_i S_{iz}\,\sigma \,c^\dagger_{\sigma}({\bf r}_i)  c^ \pdag_{\sigma}({\bf r}_i)~,
\end{align}
where $S_{iz}=\pm|{\bf S}_i|$ is the $\hat{z}$ component of the spin of impurity $i$, and $J_i$ is the exchange coupling between this impurity and the electrons in the superconductor. This model comes with the Debye frequency $\omega_D$ as a natural high-energy cutoff.  

If the renormalization of the superconducting gap is small, $|\delta \Delta({\bf r})| = |\Delta({\bf r})-\Delta_0| \ll \Delta_0$, one can approximate $\Delta({\bf r})$ by evaluating the right-hand side of the gap equation for an unrenormalized gap,
\begin{align}
\Delta({\bf r}) \approx -g\,\langle c_\downarrow({\bf r}) c_\uparrow({\bf r})\rangle_{H'}\label{eq:gap_eq}
\end{align}
with $H'=H_{0}+H_{\rm imp}$. This is the approach we employ in the remainder of the analytical calculation. When $|\delta\Delta({\bf r})|$ becomes of order $\Delta_0$, this approximation ceases to be valid, and we make use of numerical simulations. The tight-binding Hamiltonian $\bar H$ is defined as
\begin{align}
&\bar H=-t \sum_{<i,i'>}\sum_{\sigma=\pm1} c_{i\sigma}^\dagger c_{i'\sigma} \\
&  + \sum_{i=1}^{N_x  \cdot N_y} \sum_{\sigma=\pm1}[ \Delta_i \sigma c_{i  \sigma} c_{i \bar \sigma}- (\mu - 4t +\bar J_i \sigma) c_{i\sigma}^\dagger c_{i\sigma}
 ]+\text{H.c.},\nonumber
\end{align}
where $c_{i\sigma}$ is the annihilation operator acting on an electron with spin $\sigma$ at lattice site $i$, and the first sum runs over neighboring sites $i$ and $i'$ located in a two-dimensional square  lattice of size $N_x \times N_y$ with lattice constant $a$. The chemical potential $\mu$ is taken from the bottom of the energy band, and the local order parameter  $\Delta_i$ is determined self-consistently in an iterative procedure for fixed value of the exchange coupling $\bar J_i$ at the site $i$ starting from the uniform superconducting order parameter $\Delta_0$ until convergence is reached.~\cite{flatte_97,salkola_97,flatte_00,morr_06,balatsky_06}

\section{Analytical T-matrix approach}\label{sec:t_matrix}
We  begin with calculating the full imaginary time Nambu Green's function $\mathcal{G}({\bf r},\textbf{r}',\tau,\tau')=-\langle T_\tau\Psi(\textbf{r}',\tau') \Psi^\dagger(\textbf{r},\tau) \rangle_{H'}$ using the `bare' Green's function $\mathcal{G}_0({\bf r},\textbf{r}',\tau,\tau')=-\langle T_\tau  \Psi(\textbf{r}',\tau')\Psi^\dagger(\textbf{r},\tau)\rangle_{H_0}$, and where the Nambu spinor is defined as $\Psi_{\bf k} = (c_{ {\bf k}\uparrow}^\pdag,c_{ -{\bf k}\downarrow}^\dagger)^T$.
For equal positions $\bf{r}=\bf{r}'$, a Fourier transformation from imaginary time to Matsubara frequencies $\omega_n$ yields
\begin{align}
\mathcal{G}_0({\bf r},\textbf{r},\omega_n)&=-\frac{\pi\nu_F}{\sqrt{\omega_n^2+\Delta_0^2}}\,\left(i\omega_n\tau_0+\Delta_0\tau_x\right)~,
\end{align}
with $\nu_F$ being the density of states (per spin) at the Fermi energy, and with the Pauli matrices $\tau_i$ acting in Nambu space ($\tau_0 = \mathds{1}_{2\times2}$), while we obtain
\begin{align}
&\mathcal{G}_0({\bf r},\textbf{r}',\omega_n)=-\frac{\pi\nu_F}{\sqrt{\omega_n^2+\Delta_0^2}}\frac{e^{-\sqrt{\omega_n^2+\Delta_0^2}\delta r/v_F}}{k_F \delta r}\label{eq:g0rrp}\\
&\times\Bigl(\sin(k_F \delta r)\left(i\omega_n\tau_0+\Delta_0\tau_x\right)+\cos(k_F \delta r)\sqrt{\omega_n^2+\Delta_0^2}\,\tau_z\Bigr)\nonumber~,
\end{align}
at distances $\delta r = |\bf{r}-\bf{r}'|$ larger than the Fermi wavelength ($k_F$ is the Fermi momentum). Using the imaginary time-dependent Dyson equation, the full Green's functions can be expressed as 
\begin{align}
\mathcal{G}({\bf r},\textbf{r}',\omega_n) &= \mathcal{G}_0({\bf r},\textbf{r}',\omega_n)\nonumber\\
&+\sum_{i,j}\mathcal{G}_0({\bf r},\textbf{r}_i,\omega_n)\,{T}_{n}^{ij}\,\mathcal{G}_0({\bf r}_j,\textbf{r}',\omega_n)~,\label{eq:tmatrix}
\end{align}
where the $T$-matrix ${T}_{n} = {V}\,\left[\mathds{1}_{4\times4}-{G}_{0,n}{V}\right]^{-1}$ involves the $4\times4$ -matrix ${G}_{0,n}$ with entries ${G}_{0,n}^{ij} =\mathcal{G}_0({\bf r}_i,\textbf{r}_j,\omega_n)$, and ${V} = \text{diag}(J_1 S_{1z},J_1 S_{1z},J_2 S_{2z},J_2 S_{2z})$.

\section{Single impurity physics}\label{sec:single_imp}
When the impurities are far apart from each other, $(k_Fr_{12})^{-1}\ll1$, the gap near a given impurity is predominantly renormalized by scattering processes off this impurity. \cite{rusinov_69,schlottmann_76} We can then for instance focus on impurity $1$.
Next to this impurity, and using $\alpha_i = \pi \nu_F J_i S_{iz}$, the dominant contribution to the gap renormalization at $\textbf{r}_1$ reads\cite{rus_schlott_comm} 
\begin{align}
&\delta\Delta^{(1)}(\textbf{r}_1) = \frac{g\pi|\alpha_1|\Delta_0\nu_F}{(1+\alpha_1^2)^2}\label{eq:deltadeltasmall}\\
&-\frac{g\alpha_1^2\Delta_0}{2}\int_{-\infty}^{\infty} d\omega\,\frac{\nu_F}{\sqrt{\omega^2+\Delta_0^2}}\frac{\omega^2(3+\alpha_1^2)+4\Delta_0^2}{\omega^2(1+\alpha_1^2)^2+4\alpha_1^2\Delta_0^2}\nonumber~.
\end{align}
A Gaussian high-energy cutoff at the scale $\omega_D$ regularizes the logarithmic UV divergence of this integral. For small $|\alpha_1|$, we can furthermore expand this expression to second order in $\alpha_1$. Since the self-consistent equation \eqref{eq:gap_eq} yields a bare gap of $\Delta_0\approx2\omega_D\,e^{-1/g\nu_F}$, we find
\begin{align}
\delta\Delta(\textbf{r}_1) &\approx-3\alpha_1^2\Delta_0~.\label{eq:deltadeltasmall3}
\end{align}
For a more strongly bound Shiba state, $|\alpha_1| \to 1$, our analytical calculation hints at a very strong suppression of the gap at the positions of the impurities, $|\delta \Delta({\bf{r}}_i)|\sim\Delta_0$, but the non-selfconsistent approach cannot make any quantitative predictions in this regime. Numerically, on the other hand, we can readily access the regime  $|\alpha_1| \geq 1$, see Fig.~\ref{Two_impurity}. For small values of $\bar J_1$, the gap under a (single) impurity is suppressed quadratically in $\bar J_1$ in agreement  with Eq. (\ref{eq:deltadeltasmall3}). Interestingly, at a critical value $\bar J_{c}$  corresponding to the phase transition,\cite{{flatte_97,salkola_97}} the superconducting gap under the impurity changes its sign and magnitude abruptly, giving rise to a local $\pi$ junction, see Fig.~\ref{Two_impurity}a. Also the energy of the bound state jumps from zero to a finite negative value. This feature results from self-consistency in determining $\Delta({\bf r}) $ and cannot be captured analytically in the approach we used above. The stronger $\bar J_1$ becomes, the smaller $\Delta_1$ under the impurity gets, until the superconducting gap is eventualy totally suppressed. As a result, the internal S-S' junction evolves into an S-N junction.

\section{Shiba to Andreev state transition}\label{sec:shiba_andreev}
\begin{figure}[!t]
\includegraphics[width=0.9\linewidth]{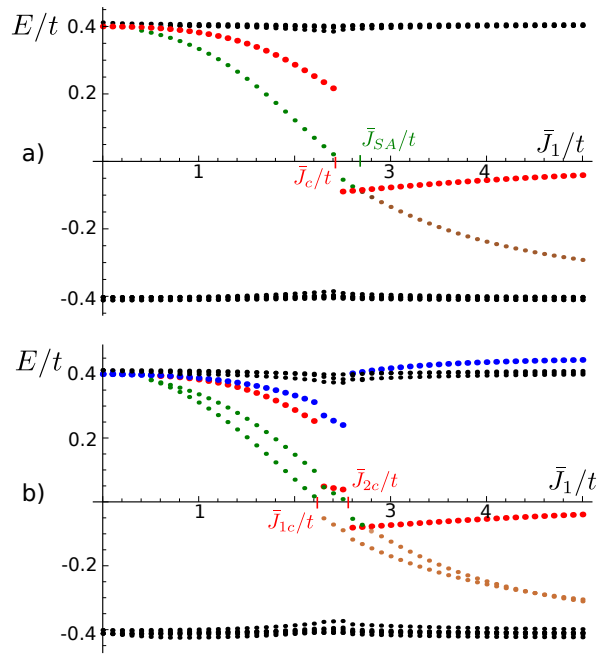}
\caption{The energy spectrum (black, green, and brown dots) and the size of the superconducting gap $\Delta_1$ directly under the impurity (red dots) for a square lattice of size $17a \times 13 a$  with (a) a single impurity  (b) two identical impurities ($\bar J_1=\bar J_2$) as function of exchange $\bar J_1/t$ found numerically.  
$\Delta_1$ gets suppressed with increasing $\bar J_1$ and  an effective S-S' junction builds up.
After the phase transition at the critical value $\bar J_{1c}$,  $\Delta_1$ changes its sign giving rise to a $\pi$-junction. Sub-gap Shiba states (green dots) evolve into supra-gap Andreev states (brown dots) at values of $\bar J_{SA}$ where  the bound state energy coincides with $\Delta_1$.
(b) For two impurities, separated by a distance $x_{12}/a=4$, the hybridization between the two bound states lifts their degeneracy. There are thus two phase transitions, one at $\bar J_{1c}$ and one at $\bar J_{2c}$, where these states become occupied, accompanied by jumps in bound state energies and, hence, in $\Delta_1$ and $\Delta_2$. There are also two distinct Shiba to Andreev state transitions. The gap between the impurities (blue dots) is suppressed for small $\bar J_1$, when the two bound states have finite overlap, but restores at larger $\bar J_1$, see also Fig.~\ref{Cross_section}. For corresponding wavefunction plots, see Fig.~\ref{Two_impurity_numerical}. The  chosen parameters are   $\Delta_0/t=0.4$ and $\mu/t=0.9$. }
\label{Two_impurity}
\end{figure}

\begin{figure}[!b]
\includegraphics[width=\linewidth]{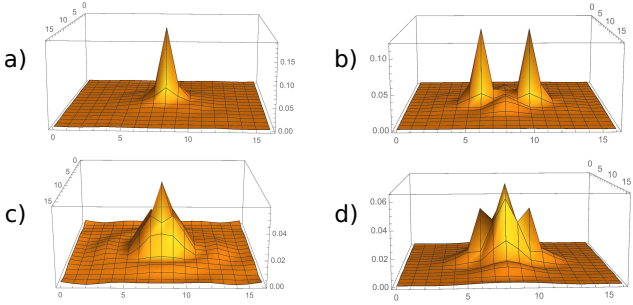}
\caption{The  wavefunctions $|\psi |^2$ of the bound states in the case of a single impurity (a,c) and of two identical impurities (b,d), $J_1=J_2=J$ found in the tight-binding model on a lattice of  size $17a \times 17 a$ by self-consistent numerical diagonalization. For two impurities, separated by a distance $x_{12}/a=4$, the hybridization between the two bound states lifts their degeneracy. Before the phase transition with $J/t=2$ (a,b), the Shiba bound state is well localized directly under the corresponding impurity. The Andreev bound state  occurring after the phase transition at stronger values of the exchange coupling, $J/t=4$ (c,d), is delocalized over the entire region where the superconducting gap is suppressed, with a maximum wavefunction amplitude between the impurities, see (d).The  chosen parameters are   $\Delta_0/t=0.4$ and $\mu/t=0.9$. }
\label{Two_impurity_numerical}
\end{figure}

After the phase transition, the modulus of the bound state energy increases, whereas the superconducting gap $|\Delta_1|$ decreases as functions of $\bar J_1$. This means that the energy of the bound state lies outside the gap for  $\bar J_1 >\bar J_{SA}$. We refer to this state as an Andreev bound state localized in an effective S-S' junction. The transition is clearly visible in our numerical results plotted in Fig.~\ref{Two_impurity}a. A key signature distinguishing Shiba and Andreev bound states is their degree of localization. This is demonstrated in Fig.~\ref{Two_impurity_numerical}, which shows the numerically obtained bound state wave functions. Indeed, the Shiba states are more localized under the impurity, see Fig.~\ref{Two_impurity_numerical}a. In contrast, the Andreev bound states are delocalized directly under the impurity as their energies are higher than the superconducting gap in this region, see Fig.~\ref{Two_impurity_numerical}b. This difference in the characteristics of the bound state wavefunctions  can be checked experimentally with STM techniques. We note that the filling of the Shiba state does not necessarily coincide with the Shiba to Andreev transition, see again Fig.~\ref{Two_impurity}a. As a consequence, there exists a finite range of exchange couplings with a filled but localized Shiba state, which delocalizes at the Shiba to Andreev transition. Our results agree qualitatively with earlier numerical studies of a single impurity,~\cite{salkola_97,balatsky_06} which, however, do not address the  Shiba to Andreev state transition.

\begin{figure}[!b]
\includegraphics[width=0.8\linewidth]{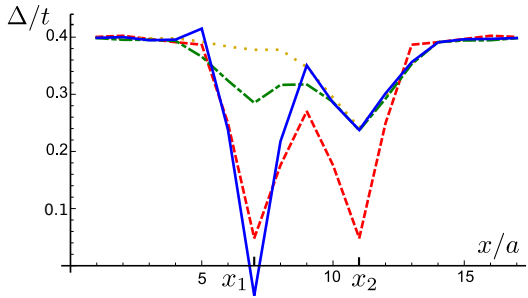}
\caption{The cross-section of the spatial profile of the  gap $\Delta$ for two impurities located on the $x$ axis for fixed  $\bar J_2/t=2.3$ and varying  $\bar J_1$:
$\bar J_1/t=1$ (yellow dotted  line), $\bar J_1/t=2$ (green dotted dashed line),  $\bar J_1/t=2.3$ (red  dashed line), $\bar J_1/t=4$  (blue solid line). The gap under the first impurity is increasingly suppressed with increasing exchange coupling: After the phase transition $\bar J_1 > \bar J_{c}$,  $\pi$-junctions (negative gap value) under the impurity arise accompanied by restoration of the gap between the two impurities. The gap under the second impurity is suppressed the most when  $\bar J_1\approx  \bar J_2$ since then the two bound states hybridize strongly. For wavefunction plots, see Fig.~\ref{Two_impurity_numerical}.
All parameters are the same as in Fig.~\ref{Two_impurity}b.}
\label{Cross_section}
\end{figure}

To obtain further insight into the sub-gap Shiba to supra-gap Andreev state transition, we study the case of a single impurity with a classical spin $S = |\textbf{S}|$ at position ${\bf r}=0$. We use the simple gap renormalization  $-a^3\Delta'\delta({{0}})$ to mimic a suppression of the gap from $\Delta_0$ to $\Delta_0-\Delta'$ within a region of volume $a^3$ around the impurity (note that we require $\Delta'>0$ since there cannot be a bound state with energy larger than the gap if the latter is locally enhanced).
Inverting the Hamiltonian,\cite{pientka_13} we obtain an equation for  the wave function at the impurity site, $\psi(0)$, in terms of the bound state energy $E$,
\begin{align}
\psi( 0)=\int\frac{d\textbf p}{(2\pi)^3} \frac{(E+\xi_{\textbf p}\tau_z+\Delta_0\tau_x)(\mp J S -a^3\Delta'\tau_x)}{E^2-\xi_{\textbf p}-\Delta_0^2}\psi(0),
\end{align}
where $\xi_{\textbf p}=p^2/2m -\mu$, and $\mp$ corresponds to the parallel and antiparallel orientation of the bound state spin as compared to the impurity spin, respectively. This equation can be rewritten as
\begin{align}\left[ 1- \frac{\tau_x(\alpha' E\pm\alpha\Delta_0)+(\alpha'\Delta_0\pm \alpha E)}{\sqrt{\Delta_0^2-E^2}}\right]\psi( 0)=0\,,
\label{an_H}
\end{align}
where $\alpha'=\nu_F\pi a^3 \Delta'$, and $\alpha = \pi \nu_F J S$. We find that bound state energies $E$ within the bulk gap $\Delta_0$ satisfying Eq.~\eqref{an_H} are given by
\begin{align}
\frac{E}{\Delta_0}=\tau\frac{ 1-w^2}{1+w^2}\,,
\end{align}
where $\tau$ is the eigenvalue of $\tau_x$, and $w=\tau\alpha'\pm\alpha$. Plugging this expression for $E$ back into Eq.~\eqref{an_H}, we find that a necessary condition is $|w|-\tau w=0$. Because consistency with BCS theory requires $\Delta_0,~\Delta'\ll E_F$, and since we are primarily interested in the regime where $|\alpha|$ is of order unity, we find that $|\alpha|\gg|\alpha'|$. Therefore, $\tau$ and $\pm \alpha$ must be of the same sign, and 
there are two solutions with opposite spin like for the uniform case.\cite{shiba_68} 
Defining as above the bound state as a Shiba state when its energy is within the renormalized gap, $|E|<|\Delta_0-\Delta'|$, and as an Andreev state when its energy is between the renormalized gap and the bulk gap, $|\Delta_0-\Delta'|<|E|<\Delta_0$, the critical value of the exchange coupling for the transition from a Shiba to an Andreev state reads 
\begin{align}
|\pi\nu_F J_{\textrm{SA}}S|=-\alpha'+\sqrt{\frac{\Delta'}{2\Delta_0-\Delta'}}\approx\sqrt{\frac{\Delta'}{2\Delta_0-\Delta'}}\,\label{eq:jsa}
\end{align}
for $\Delta'>\Delta_0$, $|\alpha|+\alpha'>1$ (as in the numerics), and for $\Delta'<\Delta_0$, $|\alpha|+\alpha'<1$, while otherwise the fraction under the square root has to be inverted.

\section{Two impurity physics}\label{sec:two_imp}
\begin{figure}
  \centering
\includegraphics[width=0.7\columnwidth]{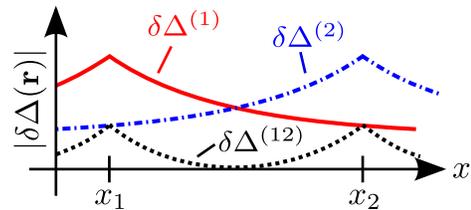}\\
  \caption{Sketch of the renormalization of the superconducting gap, $\delta \Delta({\bf r}) = \Delta({\bf r})-\Delta_0$,
   in the presence of two spin-impurities localized at positions ${\bf r}_1$ and ${\bf r}_2$ along the $\hat{x}$ axis connecting the two impurities. 
   The sketch shows the envelopes of three different contributions to the gap renormalization (see text).}
  \label{fig:gap_renorm}
\end{figure}

When the impurities are further apart than the Fermi wavelength, $r_{12} = |\textbf{r}_1-\textbf{r}_2| \gg k_F^{-1}$, we can expand the gap renormalization in orders of $(k_F r_{12})^{-1}$. To second order, we find 
\begin{align}
&\mathcal{G}({\bf r_1},\textbf{r}_1,\omega_n) \approx {G}_{0,n}^{11}+ {G}_{0,n}^{11} {T}_{n}^{(1)}{G}_{0,n}^{11}\nonumber\\
&+ \left(1+{G}_{0,n}^{11}{T}_{n}^{(1)}\right){G}_{0,n}^{12}{T}_{n}^{(2)}{G}_{0,n}^{21}\left(1+{T}_{n}^{(1)}{G}_{0,n}^{11}\right)\label{eq:two_imp_order_kfr},
\end{align}
where ${T}_{n}^{(i)} = J_i S_{iz} \left[\mathds{1}_{2\times2}-J_i S_{iz} \,{G}_{0,n}^{ii}\right]^{-1}$ is the $T$-matrix for scattering off impurity $i=1,2$. The terms ${G}_{0,n}^{1i}{T}_{n}^{(i)}{G}_{0,n}^{i1}$ give rise to the single impurity renormalizations of the order parameter, which we denote as $\delta \Delta^{(i)}$, while the other corrections to the bare Green's function result in inter-impurity gap renormalizations $\delta\Delta^{(12)}$. These different contributions are depicted in Fig.~ \ref{fig:gap_renorm}. The inter-impurity renormalizations have both terms that depend on the relative orientation of the two spins (parallel for $\alpha_1\alpha_2 > 0$, antiparallel for $\alpha_1\alpha_2 < 0$), as well as orientation-independent contributions. All of these are of the order $(k_Fr_{12})^{-2}$. 
Most interesting is the orientation-dependent contribution evaluated at the site of one of the impurities. As an example, $\delta\Delta^{(12)}_{\pm}(\textbf{r}_{1})$ reads at zero temperature
\begin{align}
&\delta\Delta^{(12)}_{\pm}(\textbf{r}_{1}) =\mp |\alpha_1\alpha_2|\,g\Delta_0\nu_F \int d{\omega}
\frac{e^{-\frac{2{r_{12}}}{v_F}
\sqrt{\omega^2+\Delta_0^2}}}{k_F^2r_{12}^2}\nonumber\\
\times& \frac{(\omega^2+\Delta_0^2)^{3/2}}{\mathcal{A}_{1}^2\mathcal{A}_{2}}\Bigl\{(1-\alpha_2^2)[(1+\alpha_1^2)^2\omega^2-(1-\alpha_1^2)^2\Delta_0^2]\nonumber\\
&-2(1-\alpha_1^4)(1+\alpha_2^2)\omega^2\cos(2k_Fr_{12})\Bigr\},\label{eq:renorm_12}
\end{align}
with $\mathcal{A}_{i}=(1-\alpha_i^2)^2\Delta_0^2+(1+\alpha_i^2)^2\omega^2$.  The upper (lower) sign applies for parallel (antiparallel) impurity spin alignment.

Besides addressing the gap renormalization at the sites of the impurities, it is also interesting to analyze the scaling of the different contributions to the gap renormalization as a function of the inter-impurity distance in their middle, that is at $\textbf{R}=(\textbf{r}_1+\textbf{r}_2)/2$. The single impurity contributions are found to scale as 
\begin{align}
\delta\Delta^{(i)}(\textbf{R}) \sim \Delta_0\,e^{-r_{12}/\xi}\, (k_Fr_{12})^{-2}~,
\end{align}
where $\xi$ is the superconducting coherence length. This scaling has a simple geometrical interpretation: to leading order in $(k_F r_{12})^{-1} \ll1$, the anomalous Green's function (which determines the gap) is renormalized by the electron traveling from $\textbf{R}$ to impurity $i$ at $\textbf{r}_i$, scattering there, and coming back to $\textbf{R}$. Since a trip to, and back from, the impurity involves a propagator proportional to $(k_Fr_{12})^{-1}$, see Eq.~\eqref{eq:g0rrp}, the gap renormalization deriving from the single impurity scattering processes scales as $(k_Fr_{12})^{-2}$. The total distance covered during these processes, precisely equal to $r_{12}$, determines the argument of the exponential. The leading order inter-impurity terms, on the other hand, are found to scale as 
\begin{align}
\delta\Delta^{(12)} (\textbf{R}) \sim \Delta_0\,e^{-2r_{12}/\xi}\,(k_Fr_{12})^{-3}~,
\end{align}
since they stem from an electron traveling first to impurity one, then to impurity two, and then coming back. This involves three trips, and a total distance of $2r_{12}$. The inter-impurity renormalization $\delta\Delta^{(12)}$ is thus suppressed by an additional power of $k_F r_{12}\gg1$ in between the impurities as compared to its value at one of the impurities, see Eq.~\eqref{eq:renorm_12}. Consequently, the inter-impurity gap renormalization $\delta\Delta^{(12)}$ can be modeled as a function with well-defined peaks close to the two impurities, as shown in Fig.~\ref{fig:gap_renorm}. We also confirm numerically that the renormalizations of the superconducting gap is the largest directly under the impurity, while, depending on parameters, $\Delta(\bf r)$ could  even get {\it larger} than its initial value around the impurities, see Fig.~\ref{Cross_section}.  We finally note that the regime $k_Fr_{12}\sim1$ is qualitatively similar to the regime $k_Fr_{12}> 1$. This is illustrated by Fig.~\ref{Two_impurity}b, which is not far from this point.

With increasing exchange strengths, a similar transition as before between Shiba and Andreev states can also be observed in the case of two impurities, see Fig.~\ref{Two_impurity}b. 
When the two impurities are close and with comparable exchange strengths,  the bound states overlap and their degeneracy gets lifted by hybridization,~\cite{flatte_00,zarand_14} which in turn gives  rise to two separate quantum phase transitions in $\Delta$, one at $\bar J_{1c}$ and one at $\bar J_{2c}$.\cite{morr_06}  Similarly, we find now two distinct Shiba to Andreev state transitions around these values of exchange strengths, see Fig.~\ref{Two_impurity}b. Some illustrative plots of the corresponding wavefunctions are given in Fig.~\ref{Two_impurity_numerical}. The spectroscopic signatures of the Shiba to Andreev transition are now even more drastic than for the single impurity case. As shown in Fig.~\ref{Two_impurity_numerical}b, the bound state wavefunction of two strongly localized, but hybridized Shiba states is strongly suppressed between the impurities. In stark contrast, the wavefunction of the Andreev bound state spreads between the impurities, and is even maximal between the impurities for the chosen aprameters, see Fig.~\ref{Two_impurity_numerical}d.

In the case of small impurity distances $k_Fr_{12}\to 0$, where $\mathcal{G}_0({\bf r}_1,\textbf{r}_2,\omega_n) \to \mathcal{G}_0({\bf r}_1,\textbf{r}_1,\omega_n)=\mathcal{G}_0({\bf r}_2,\textbf{r}_2,\omega_n)$. Keeping all orders of $(k_F r_{12})^{-1}\gg1$ in Eq.~\eqref{eq:tmatrix}, we find that the gap renormalization is in this limit given by
\begin{align}
\delta\Delta(\textbf{r}_{1}) = \delta\Delta(\textbf{r}_{2}) &=\left.\delta\Delta^{(1)}(\textbf{r}_1)\right|_{\alpha_1\to\alpha=\alpha_1+\alpha_2},
\end{align}
where $\delta\Delta^{(1)}(\textbf{r}_1)$ is given in Eq.~\eqref{eq:deltadeltasmall}. Quite naturally, the gap renormalization resulting from two very close classical impurities with $\alpha_1$ and $\alpha_2$ is thus equal to the renormalization of a single impurity with $\alpha=\alpha_1+\alpha_2$. Provided that $J_1$ and $J_2$ have the same sign, the superconducting gap is reduced the least if the two spins are aligned antiparallel for short distances. As a final remark, we note that our non-selfconsistent analytical approach is also able to tackle the regime of intermediate impurity distances $k_Fr_{12}\sim1$. In this regime, we find some lengthy equations that simply connect the limits of small and large impurity distances in a smooth fashion. As an illustration, we refer the reader to the numerical plots in Fig.~\ref{Two_impurity}b. These are not far from the regime $k_Fr_{12}\sim1$, and even treat the gap self-consistently.

\section{Conclusions}\label{sec:concl}
Analyzing two classical spins in a superconductor, we obtained the gap renormalization analytically for weakly and numerically for strongly coupled bound states.
In addition, we found a transition for  Shiba  to  Andreev states which is accompanied by two subsequent quantum phase transitions as the exchange
coupling is varied.

All these predictions lead to dramatic spectral changes. In order to observe these effects by STM tecniques,\cite{yazdani_97,ji_08,nadj_14,meyer_15} it would be preferable to control experimentally the exchange interaction parametrized by 
$\alpha$. This can be realized for example by considering a 2D electron 
gas (with immersed magnetic impurities) deposited on top of a bulk 
superconductor such that the SC gap is now a proximity gap. We propose 
to  bring an STM tip on top of a magnetic impurity in order to locally 
control the impurity density of states without affecting its  spin and 
therefore to continuously monitor the parameter $\alpha$.

A number of additional directions can be envisioned in future work. Instead of investigating the physics associated with large impurity spins, for which a classical treatment is appropriate, one could generalize our work the limit of small impurity spins, for which Kondo physics becomes relevant at low temperatures. With two close-by impurities, this could for instance lead to interesting multi-stage screening effects. Another promising generalization is offered by the use of more exotic superconductors, such as $d$-wave materials, where impurities induce virtual bound states \cite{balatsky_06}.

\acknowledgements
We thank L. Glazman and S. Gangadharaiah for valuable discussions. This work has been supported by the Swiss NF, NCCR QSIT, by the DFG through GRK 1621 and
SFB 1143, and by the French Agence Nationale de la Recherche through the ANR contracts Dymesys and Mistral.



\begin{thebibliography}{99}
\bibitem{rus_schlott_comm}
The (non-selfconsistent) single impurity gap renormalizations reported previously~\cite{rusinov_69, schlottmann_76} are incomplete, the correct result is given in Eq.~\eqref{eq:deltadeltasmall}.
\end{thebibliography}
\end{document}